\documentclass[preprint,prb,showpacs,amsmath,amssymb]{revtex4}

\usepackage{graphicx}
\usepackage{dcolumn}
\usepackage{bm}
\usepackage{latexsym}

\begin{document}

\title{Fermi-Bose Correspondence and Bose-Einstein Condensation in The Two-Dimensional Ideal Gas}

\author{A. Swarup}%
\email[Email:]{amarendra.swarup@ic.ac.uk}%
\affiliation{Theoretical Physics, Imperial College, South Kensington, London UK SW7 2AZ}%

\author{B. Cowan}%
\affiliation{Department of Physics, Royal Holloway, University of
London, Egham, Surrey, UK TW20 OTB}%

\date{\today}%

\begin{abstract}
The ideal uniform two-dimensional (2D) Fermi and Bose gases are
considered both in the thermodynamic limit and the finite case. We
derive May's Theorem, viz. the correspondence between the internal
energies of the Fermi and Bose gases in the thermodynamic limit.
This results in both gases having the same heat capacity. However,
as we shall show, the thermodynamic limit is never truly reached in two dimensions
and so it is essential to consider finite-size effects. We show
in an elementary manner that for the finite 2D Bose gas, a
pseudo-Bose-Einstein condensate forms at low temperatures,
incompatible with May's Theorem. The two gases now have different
heat capacities, dependent on the system size and tending to the
same expression in the thermodynamic limit.
\end{abstract}


\maketitle

\section{\label{sec:intro}Introduction}

Since their inception into the theoretical framework of physics,
much work has been carried out on both the Fermi-Dirac as well as
the Bose-Einstein distributions \cite{huang87,pathria96}. Much of
this focuses on the behaviour of fermions and bosons at low
temperatures, where the deviations in their behaviour arising from
the distributions are significant and quantum effects dominate.
The advent of comprehensive experimental techniques for lowering
the temperature of systems to previously theoretical values over
the last few decades has contributed significantly to this and
revealed a wealth of unusual behaviour at extremely low
temperatures, in particular the unusual phenomenon of
Bose-Einstein condensation - a macroscopic occupation of the
ground state of a gas of bosons at finite temperature.

For most of this time, however, there has been only a cursory and
passing interest in \textit{low-dimensional} Fermi and Bose
systems. This has been mainly because they have been viewed as
little more than abstract theoretical and mathematical constructs.
Moreover, it was felt that there was little behaviour of interest
to be found in these lower dimensions. It is only over the last
couple of decades that there has been a growing interest in the
properties of gases obeying the Fermi and Bose distributions in
lower dimensions. This has been driven by firstly, the growing
practical importance of low-dimensional systems, such as for
example in computers, as well as new experimental techniques (such
as, for example, 2D adsorbed systems, in particular, 2D adsorbed
helium films \cite{lusher91,siqueira93,casey98,ray98}) that raise
the possibility of being able to trap and study low-dimensional
quantum systems. Typical systems being studied include superfluid
and superconducting films, quantum Hall and related
two-dimensional electron gases and low-dimensional trapped Bose
gases. Further, the renewed theoretical interest in
low-dimensional physics over the last couple of decades has led to
a growing realisation that these systems possess equally
fascinating and unique quantum behaviour of their own (see, for
example, Ref. \onlinecite{schakel98}). In particular, recent
observations of exotic quantum behavior, such as Bose-Einstein
condensation (BEC) \cite{anderson95,bradley97,davis95} and Fermi
degeneracy \cite{demarco99}, in three-dimensional systems have led
to a growing interest in lower-dimensional systems.

One special theoretical interest in the 2D ideal gas is due to the
fact that we can obtain completely analytic expressions for the
chemical potential $\mu$ and other thermodynamic properties. This
arises, in part, from the energy-independence of the 2D density of
states within the semi-classical approximation. Furthermore, in
two dimensions, there is an unusual correspondence between the
Bose and Fermi cases, first noted by May \cite{may64} and
subsequently, explored further in later papers
\cite{mullin2003,toda83,aldrovandi92,viefers95,lee97}.

In his seminal paper \cite{may64}, May noted the unexpected result
that the thermal capacities of the ideal 2D Bose and Fermi gases
are identical in the thermodynamic limit, i.e. their internal
energies differ only by a constant temperature-independent amount.
This only holds in the thermodynamic limit, or equivalently, the
semi-classical approximation that the energy level spacing is
negligible compared with the temperature \textit{T}, allowing us
to replace the discrete (quantum) structure by a continuum.
Real-world systems, however, are always finite, typically of the
order of $10^{3} - 10^{6}$ particles for trapped gases, and more
for adsorbed systems. Moreover, as we shall see later, the
chemical potential $\mu$ has a logarithmic dependence on the total
number of particles $N$ in the system. Thus, \textit{the
thermodynamic limit is never truly reached in two dimensions} and
is an abstract theoretical construct, making it essential to
consider finite-size effects. These may be experimentally
significant, and lead to deviations from May's Theorem.

In the following sections, we shall consider the behaviour of the
uniform 2D ideal Bose and Fermi gases. In the next
section, we shall briefly outline the gases in the
thermodynamic limit and derive May's Theorem before examining
the Bose and Fermi gases for a finite number of
particles in Section \ref{The Finite Gas}. Though there is no BEC in the thermodynamic limit, we
will demonstrate that there is a pseudo-condensation at very low
temperatures for the finite 2D Bose gas. This is then
used to show in Section \ref{Heat capacities} that the finite-size
effects lead to differing heat capacities for the Bose and Fermi
gases, and a breakdown in May's Theorem for gases with a finite
number of particles. Further, the inadequacy of heat capacities as
a means of determining phase transitions is discussed.

\section{\label{Background}The Thermodynamic Limit}

As the 2D ideal gas and May's Theorem have already been examined at length before in the thermodynamic limit\cite{may64,mullin2003,toda83,aldrovandi92,viefers95,lee97}, we shall simply draw on earlier approaches to provide a brief overview. Consider $N$ free and identical particles of mass $m$ in a 2D
\textit{box} of \textquoteleft volume\textquoteright  $A$, with
rigid-box boundary conditions. The energy values for an arbitrary
particle then are \begin{equation}
\varepsilon_{i}=\frac{\hbar^{2}k^{2}}{2m}, %
\qquad %
k^2=\frac{\pi^{2}}{A}\left(n^{2}_{x}+n^{2}_{y}\right)%
\label{eq:one},
\end{equation} where $k$ is the momentum characterising the energy
states $\varepsilon_{i}$, $i,n_x,n_y=1,2,3\ldots$
and $(n_x,n_y)$ are points in the positive quadrant of a unit
square lattice. In the thermodynamic limit, the ground state
energy tends to 0. For a finite system, however, due to the
Heisenberg uncertainty principle, this is always greater than 0, as $\left(n_{x},n_{y}\right)=
\left(1,1\right)$. So, for simplicity, we shift the energy scale,
taking the ground state energy as 0, \begin{equation}
\varepsilon_{i}=\frac{\hbar^{2}k^2}{2m}-\varepsilon_{0}= \frac{\pi^2\hbar^2}{2mA}\left(n^2_x+n^2_y-2\right)%
\label{eq:two}.
\end{equation}
Assuming that the system is in thermal equilibrium at
temperature $T$ and chemical potential $\mu$, then the mean number of
particles $n\left(\varepsilon_{i}\right)$ at a given energy
$\varepsilon_{i}$ is given by \begin{eqnarray}
n\left(\varepsilon_{i}\right)=\frac{1} {e^{\left(\varepsilon_i-\mu\right)/{k_{B}T}}\pm 1}%
\label{eq:four}.
\end{eqnarray} Here and later, the upper (lower) signs are for fermions (bosons). The total number of particles \textit{N} and the internal energy \textit{E} of the system are now given by
\begin{eqnarray}
N=\sum_{i=1}^{\infty}n\left(\varepsilon_{i}\right)%
\label{eq:five},
\\
E=\sum_{i=1}^{\infty}\varepsilon_i n\left(\varepsilon_{i}\right)%
\label{eq:six}.
\end{eqnarray}
where the sum is taken over the single particle energy states. In
the semiclassical approximation, we assume that the spacing between the energy levels of the particle states is
negligible. This allows us to replace the sum with an
integral over energy states using the density of states
$g(\varepsilon)$ and the spin degeneracy $\alpha$, \begin{eqnarray}
f=\alpha\int_{0}^{\infty}f\left(\varepsilon\right)g \left(\varepsilon\right)n\left(\varepsilon\right)d\varepsilon%
\label{eq:eight},
\end{eqnarray} where $f\left(\varepsilon\right)$ is a general single-particle function of
energy. However, this approximation is only exact in the thermodynamic limit when the system size is infinite, and at very
high temperatures $T$. As we shall see in Section \ref{The Finite Gas}, for
finite systems at low \textit{T}, these assumptions are invalid
and the discreteness of the energy levels cannot be ignored. In
other words, we will be at a temperature
$T$ below which the semiclassical approximation breaks down and
modifications need to be made. However, it may be possible to
compensate by modifying the semi-classical approximation to allow for finite-size effects.

We now evaluate the various thermodynamic quantities for arbitrary
spin, i.e. arbitrary $\alpha$, for both the 2D Fermi
and Bose gases in the thermodynamic limit. In 2D,
$g(\varepsilon)$ and thereby, the Fermi energy $\varepsilon_F$
are given by,
\begin{eqnarray}
g(\varepsilon)=\frac{mA}{2\pi\hbar^2},  \qquad
\varepsilon_F=\frac{2\pi\hbar^{2}N}{\alpha mA}%
\label{eq:nine}.
\end{eqnarray}
For comparison, we adopt $\varepsilon_F$ as a characteristic
energy $\varepsilon_C$ for the Bose gas also, allowing us to
express later results in a unified form. In 2D, as
$g(\varepsilon)$ is a constant, we have a simple integral for
\textit{N},
\begin{equation}
N=\frac{\alpha
mA}{2\pi\hbar^2}\int_0^{\infty}\frac{d\varepsilon} {e^{\left(\varepsilon-\mu\right)/{k_{B}T}}\pm1}%
\label{eq:ten}.
\end{equation}
This may be expressed using the Fermi-Dirac and Bose-Einstein
functions \cite{huang87,pathria96}, $f_\nu(z)$ and $g_\nu(z)$
respectively, \begin{eqnarray}
f_{\nu}\left(z\right)=\frac{1}{\Gamma\left(\nu\right)}\int_{0}^{\infty}\frac{x^{\nu-1}dx}{z^{-1}e^{x}+1}%
\label{eq:eleven}
\\
g_{\nu}\left(z\right)=\frac{1}{\Gamma\left(\nu\right)}\int_{0}^{\infty}\frac{x^{\nu-1}dx}{z^{-1}e^{x}-1}%
\label{eq:twelve},
\end{eqnarray} where $\Gamma\left(\nu\right)$ is the Euler gamma function. These can also be expressed using the
polylogarithmic functions $\textnormal{Li}_{\nu}(z)$
\cite{abram64,lewin58},
\begin{eqnarray}
f_{\nu}\left(z\right)=-\textnormal{Li}_{\nu}\left(-z\right), %
\label{eq:thirteen}
\\
g_{\nu}\left(z\right)=\textnormal{Li}_{\nu}\left(z\right)%
\label{eq:fourteen}.
\end{eqnarray}
For \textit{N} above, $\nu=1$ and $z$ is the fugacity
($z=e^{\mu/k_{B}T}$). Solving (\ref{eq:ten}) in terms of $\varepsilon_C$ and
rearranging, we can now express the fugacity $z$ and thereby, the
chemical potential $\mu$ as a completely analytic expression
\begin{equation}
z_{F,B}=e^{{\mu}/{k_{B}T}}=\mp\left(1-e^{\pm{\varepsilon_C}/{k_{B}T}}\right)%
\label{eq:seventeen},
\end{equation} where $z_{F}$ ($z_{B}$) is the fugacity for the ideal Fermi (Bose)
gas. As $T\rightarrow 0$, $z_B$ smoothly approaches 1. Denoting the
fugacity $z_{B}$ by $z_{B}=e^{-\alpha}$, as $z_{B}\rightarrow 1$,
\begin{equation}
g_{1}\left(e^{-\alpha}\right)=-\ln\left(1-e^{-\alpha}\right)\underset{\alpha\rightarrow
0}\longrightarrow-\ln\alpha%
\label{eq:eighteen}.
\end{equation}
\textit{N} diverges logarithmically as $\alpha\rightarrow 0$.
Thus, it has no upper bound, and there is no temperature below
which the ground state can be said to be macroscopically occupied
in comparison to the excited states. Thus, there is no BEC in the
thermodynamic limit. This is a qualification of the oft-paraded
statement that there is no BEC for $D\leq 2$, where $D$ is the
number of dimensions. However, as we shall see in Section \ref{The
Finite Gas}, the logarithmic divergence implies that at
sufficiently low \textit{T}, bosons still crowd into low-lying
states to give a pseudo-condensate.

The internal energy may be similarly solved,
\begin{eqnarray}
E_F=Nk_{B}T\left(\frac{k_{B}T}{\varepsilon_C}\right)f_{2}\left(e^{{\varepsilon_C}/{k_{B}T}}-1\right)%
\label{eq:twenty},
\\
E_B=Nk_{B}T\left(\frac{k_{B}T}{\varepsilon_C}\right)g_{2}\left(1-e^{{-\varepsilon_C}/{k_{B}T}}\right)%
\label{eq:twentyone}.
\end{eqnarray}
Now, using the following property of the dilogarithmic
function\cite{may64,lewin58} $\textnormal{Li}_2\left(x\right)$,
\begin{equation}
f_{2}\left(\frac{1}{x}-1\right)=g_{2}\left(1-x\right)+\frac{1}{2}\left(\ln
x\right)^2%
\label{eq:twentytwo},
\end{equation} we can easily show \begin{eqnarray}
E_F= E_B+\frac{1}{2}N\varepsilon_C%
\label{eq:twentythree}.
\end{eqnarray} i.e. for all $T$, $E_B$ differs from $E_F$ by a constant
temperature independent energy, which does not contribute to
the heat capacity. This gives May's Theorem \cite{may64}, which
states that the heat capacities of the ideal 2D Fermi
and Bose gases are identical at all \textit{T} and \textit{N},
\begin{equation}
C_{V,Bose}=\frac{\partial E_{B}}{\partial T}|_{_V}=\frac{\partial
E_{F}}{\partial T}|_{_V}=C_{V,Fermi}%
\label{eq:twentyfour}.
\end{equation}
The above result indicates that the 2D ideal Bose gas
does not undergo a phase transition, as any experimental
measurement of the heat capacity would be the same as for a
2D Fermi gas, and would lack the characteristic bump
that indicates the presence of BEC in the
3D Bose gas. However, as we shall see, this is because we have failed to consider the finite nature of any real systems.

\section{\label{The Finite Gas}The Finite Gas}

The case of the finite ideal Bose gas in two dimensions has
already been considered to varying degrees in earlier papers
\cite{osborne49,ziman53,
mills64,goble67,chester68,krueger68,irnry69,barber83,grossmann95},
and more recently, by Pathria \cite{pathria98}. Most have
commented that it may be possible to have a pseudo-condensation in
the Bose gas in two dimensions, with the transition temperature
typically going as $T_{C}\sim 1/\ln{N}$. Here, we present a
simpler and more comprehensive analysis of the problem that not
only offers far more insight into how the finite size of the
system leads to the breakdown of May's Theorem and a
pseudo-condensation in the Bose gas, but further, in Section
\ref{Heat capacities}, discusses how the heat capacities for the
Fermi and Bose gases differ for finite systems, and how such a
pseudo-condensation may be experimentally measured.

At $T=0$, the 2D Bose gas is in its ground state, and we have the
trivial case of ground state macroscopic occupation.
Extrapolating, however, a range of finite temperatures must exist
near $T=0$, which may depend on $N$, where the number of particles
in the ground state is also of $O\left(N\right)$ - a macroscopic
occupation. At very low $T$, this can be a significant fraction of
$N$, and the semi-classical approach is now flawed as $N$ is no
longer continuous. Taking macroscopic occupation of the ground
state as the defining characteristic of BEC, we consider the
ground state separately analogous to standard treatments of the 3D
Bose gas \cite{huang87,pathria96}. For finite systems, the
discrete level structure requires us to introduce a cutoff energy
at the first excited level $\varepsilon_1$. The integral is now
bounded and we can remove the logarithmic divergence. Thus, we can
have a pseudo-BEC in a finite two-dimensional system. This must be
associated with a breakdown of May's theorem at sufficiently low
$T$ for finite systems. We thus write (\ref{eq:ten}) for the Bose
gas as \begin{equation}
N=N_0+\frac{\alpha mA}{2\pi\hbar^2}\int_{\varepsilon_1}^{\infty}\frac{d\varepsilon} {e^{{\left(\varepsilon-\mu\right)}/{k_{B}T}}-1}%
\label{eq:twentyfive},
\end{equation} 

It should be noted that the above approximation is quite a simple and crude one, as it is only the leading term of the Euler-Maclaurin summation formula. However, it suits the purposes of our paper as our intention here is merely to highlight how even the simplest approximations show that there is a pseudo-condensation in the finite Bose gas. A more detailed approximation will only lead to minor modifications to the core results presented in the rest of this paper. Further, the leading correction term to the integral above in the Euler-Maclaurin expansion is of the order of $N_{1}=n\left(\varepsilon_{1}\right)=\left(e^{\left(\varepsilon_{1}-\mu\right)/k_{B}T}\right)^{-1}$. Taking the ratio $N_{0}/N_{1}$ of the number of particles in the ground state to that in the first excited state, we can see
\begin{equation}
\frac{N_{0}}{N_{1}}\approx\frac{\varepsilon_{1}-\mu}{-\mu}%
\label{eq:a}
\end{equation}
As the chemical potential $\mu$ for the 2D Bose gas is always negative and less than $0$, the above ration is always greater than 1. Further, as $T\rightarrow 0$, $\mu\rightarrow 0$ and the above expression clearly goes to $\infty$. Further, a more detailed evaluation of (\ref{eq:a}) gives us
\begin{equation}
\frac{N_{0}}{N_{1}}\approx\frac{3\pi^{2}\hbar^{2}}{2mAk_{B}T}e^{\frac{2\pi\hbar^{2}N}{\alpha mAk_{B}T}}%
\label{eq:b}
\end{equation}
This clearly rapidly goes to infinity as both the temperature $T$ as well as the number of particles $N$ increase. Moreover, for the regime in question, we are interested in very low temperatures where from the above equation (\ref{eq:b}), the ratio $N_{0}/N_{1}$ is clearly significantly large.

Thus, we are justified in making our crude approximation and denoting the number of particles $N$ by (\ref{eq:twentyfive}) above. Solving as before, we now have \begin{equation}
N-N_0=-\frac{\alpha mA k_{B}
T}{2\pi\hbar^2}\ln\left(1-z_{B}e^{{-3\pi^2\hbar^2}/{2mAk_{B} T}}\right)%
\label{eq:twentysix}
\end{equation} and $\varepsilon_1$ prevents the logarithmic divergence. This is
because the excited states cannot accommodate all the particles at
low $T$, and the excess are forced into the ground state, leading
to a pseudo-BEC. Equivalently, changing variables
$\varepsilon\rightarrow\varepsilon +\varepsilon_{1}$ and resetting
the integral from zero, we may also express (\ref{eq:twentyfive})
in terms of the Bose-Einstein functions $g_\nu\left(z\right)$ as
\begin{equation} N-N_0=\frac{\alpha mAk_{B}
T}{2\pi\hbar^2}g_{1}\left(z_{B}e^{{-3\pi^2\hbar^2}/{2mAk_{B} T}}\right)%
\label{eq:twentyseven}.
\end{equation} Thus, we now have an effective fugacity \begin{eqnarray}
z_{B}\longrightarrow z_{B}'=z_{B}e^{{-3\pi^2\hbar^2}/{2mAk_{B}
T}}\nonumber
\\
=e^{{-3\alpha\pi\varepsilon_C}/{4Nk_{B}
T}}\left(1-e^{{-\varepsilon_{C}}/{k_{B}T}}\right)%
\label{eq:twentyeight}.
\end{eqnarray} where $z_{B}$ is as before. This summarizes the finite-size
effects on the Bose gas. The extra term is dependent on the system
size and tends to 1 as $N\rightarrow\infty$. For finite $N$ and
low $T$, it is less than 1, and so, the fugacity is modified.
Thus, there is a pseudo-BEC here but none in the thermodynamic
limit. This macroscopic occupation is a purely boson phenomenon,
and so, the behaviour will be markedly different from that of the
two-dimensional Fermi gas.

For the Fermi gas, Pauli's principle forbids macroscopic
occupation of any state. Thus, the first term is not divergent.
$z_{F}$ (Eq. (\ref{eq:seventeen})) still holds, and finite-size
effects only appear for the Bose gas, where the excited states
take as many particles as possible at low $T$, with the excess
forced into the ground state. This occurs at the maximal value of
the fugacity $z_{B}=1$. Thus, we can define a characteristic
temperature $T_C$ that marks the onset of condensation where
$N_{0}\approx 0$ and $z\approx 1$,
\begin{eqnarray}
N\cong-\frac{\alpha mAk_{B}
T_{C}}{2\pi\hbar^2}\ln\left(1-e^{{-3\pi^2\hbar^2}/{2mAk_{B}
T_{C}}}\right)\nonumber
\\
\simeq\frac{Nk_{B} T_{C}}{\varepsilon_C}\ln\left(\frac{4Nk_{B}
T_{C}}{3\pi\alpha\varepsilon_C}\right)&%
\label{eq:twentynine}
\end{eqnarray}
for very low $T$. We may solve (\ref{eq:twentynine}) to give
\begin{equation}
T_C=\frac{\varepsilon_C}{k_B}\frac{1}{W \left({4N}/{3\pi\alpha}\right)}%
\label{eq:thirty},
\end{equation}
where, $W\left(x\right)$ is the Lambert function \cite{corless96}
and is the principal solution for $w$ in the equation $x=we^w$.
This is plotted in Fig. 1. It may be viewed as a generalisation to
a logarithm. Thus, our result (\ref{eq:thirty}) is a refinement of
the earlier noted result that the transition temperature behaves
like $T_{C}\sim 1/\ln{N}$. As $N\rightarrow\infty$, the reduced
characteristic temperature
${k_{B}T_{C}}/{\varepsilon_C}\rightarrow 0$. Thus, in the
thermodynamic limit, we recover the conventional result that there
is no BEC in two dimensions. The condensate fraction is
\begin{eqnarray}
\frac{N_0}{N}\cong1-\frac{k_{B}T}{\varepsilon_C}\ln \left(\frac{4Nk_{B}T}{3\pi\alpha\varepsilon_C}\right)%
\nonumber
\\
\cong 1-\frac{T}{T_C}\left[1+\frac{k_{B}T_{C}}{\varepsilon_C}\ln\left(\frac{T}{T_C}\right)\right]%
\label{eq:thirtyone}.
\end{eqnarray}
This only holds for very low temperatures $T<T_C$. As can be seen
in Fig. 2, the condensate occurs at lower and lower $T$ as the
system size $N$ increases. In the thermodynamic limit, there is no
BEC at all.

Having shown the pseudo-condensation, we now move on to consider
how the Fermi-Bose correspondence predicted by May is affected
when the gases are assumed finite.

\section{\label{Heat capacities}The Finite Gas: Internal Energy and Heat Capacity}

The internal energies may be similarly found. The first term in
(\ref{eq:six}) for the Bose gas is divergent as it is the internal
energy $E_{gs}$ associated with the anomalously large ground
state. Separating this out,
\begin{equation}
E_{B}=E_{gs}+\sum_{i=1}^{\infty}\frac{\varepsilon_i}{e^{\left(\varepsilon_i-\mu\right)/{k_{B}T}}-1}%
\label{eq:thirtytwo},
\end{equation}
As we have set $E_{gs}$ as the zero-point energy, the first term
is 0. Changing variables
$\varepsilon\rightarrow\varepsilon+\varepsilon_{1}$, and
converting to an integral, \begin{eqnarray}
E_B=\frac{N}{\varepsilon_C}\int_0^{\infty}\frac{\varepsilon+\varepsilon_1} {e^{\left(\varepsilon+\varepsilon_1-\mu\right)/{k_{B}T}}-1}d\varepsilon%
\nonumber
\\
=\frac{N}{\varepsilon_C}\left(k_{B}T\right)^{2}g_{2}\left(z_{B}'\right)+ \frac{\varepsilon_1}{\varepsilon_C}Nk_{B}Tg_{1}\left(z_{B}'\right)%
\label{eq:thirtythree}.
\end{eqnarray} where $z_{B}'$ is defined by (\ref{eq:twentyeight}). We again have
an effective fugacity $z_B'$ and additional terms, which are
negligible in the thermodynamic limit. By contrast, for the finite
Fermi gas, there is no divergence in the ground state term, and we
have the same expression for the fugacity (\ref{eq:seventeen}) as
before. The internal energies of the finite two-dimensional Fermi
and Bose gases are then given by
\begin{eqnarray}
E_F&=&Nk_{B}T\left(\frac{k_{B}T}{\varepsilon_C}\right)f_{2} \left(e^{{\varepsilon_C}/{k_{B}T}}-1\right)%
\label{eq:thirtyfour}
\\
E_B&=&\left(\frac{Nk_{B}^{2}T^{2}}{\varepsilon_C}\right)g_{2}\left(z_{B}'\right) -\frac{3\pi\alpha}{4}k_{B}T\ln\left(1-z_{B}'\right)%
\label{eq:thirtyfive}
\end{eqnarray}
Eq. (\ref{eq:twentythree}) is no longer true, as we cannot simply
express $E_F$ in terms of $E_{B}$ as before. $E_B$ now has
additional $T$-dependent terms that lead to a differing heat
capacity.

May's Theorem is thus a special limiting case for the 2D Bose and
Fermi internal energies in the thermodynamic limit. Here, the
additional factor $e^{{-3\pi\alpha\varepsilon_{C}} /{4Nk_{B}T}}$
tends to unity in the thermodynamic limit, recovering May's
Theorem. The same occurs in the high $T$ limit for a finite gas as
above $T_C$, there is no macroscopic ground state occupation.
Thus, above $T_C$, we may approximate the internal energies by
(\ref{eq:twenty}) and (\ref{eq:twentyone}) with differences
becoming marked only at very low $T$ and/or very small system size
$N$.

We now calculate the Bose and Fermi heat capacities by using the
following polylogarithmic identity \cite{pathria96,abram64},
\begin{equation}
\frac{d}{dz}Li_{n+1}\left(z\right)=\frac{1}{z}Li_{n}\left(z\right)%
\label{eq:thirtysix}
\end{equation}
The heat capacity for the finite gas is then given by
\begin{widetext}
\begin{eqnarray}
C_{V,Fermi}=\frac{Nk_{B}^2}{\varepsilon_C}\left[2Tg_{2} \left(1-e^{{-\varepsilon_C}/{k_{B}T}}\right)-\frac{\varepsilon_{C}^{2}}{k_{B}^{2}T} \frac{1}{e^{{\varepsilon_C}/{k_{B}T}}-1}\right]%
\label{eq:thirtyseven}
\\
C_{V,Bose}=
\frac{Nk_{B}^2}{\varepsilon_C}\left[2Tg_{2}\left(z_{B}'\right)+\frac{\varepsilon_{C}}{k_{B}} \frac{\ln\left(1-z_{B}'\right)}{e^{{\varepsilon_C}/{k_{B}T}}-1}\right]%
\nonumber
\\
+\frac{3\pi\alpha k_{B}}{16 N} \left[\frac{\varepsilon_C \left(3\pi\alpha z_{B}'-4Ne^{-{\varepsilon_C}/{k_{B}T} - {3\alpha\pi\varepsilon_C} /{4Nk_{B}T}}\right)}{k_{B}T\left(1-z_{B}'\right)}-8N\ln\left(1-z_{B}'\right)\right]%
\label{eq:thirtyeight}
\end{eqnarray}
\end{widetext}
The Bose heat capacity $C_{V,Bose}$ now contains several terms
dependent on both $N$ and $T$. It should also be noted that the
last two terms in the heat capacity are very much smaller than the
first two terms and become increasingly negligible as the system
size increases. At high $T$ and/or very
large $N$, the additional terms tend to zero, i.e. as
$N\rightarrow\infty$ and/or $T\rightarrow\infty$,
${\pi\alpha\varepsilon_C}/{Nk_{B}T}\rightarrow 0$, and
(\ref{eq:thirtyeight}) reduces to the standard expression in the
thermodynamic limit,
\begin{widetext}
\begin{eqnarray}
C_{V,Bose}\rightarrow
\frac{2Nk_{B}^{2}T}{\varepsilon_C}g_{2}\left(1-e^{{-\varepsilon_C}/{k_{B}T}}\right) -\frac{N\varepsilon_C}{T}\frac{1}{e^{{\varepsilon_C}/{k_{B}T}}-1}=C_{V,Fermi}%
\label{eq:thirtynine}.
\end{eqnarray}
\end{widetext}
This is one of the most important results of our paper, and shows
clearly that May's theorem arises as the thermodynamic limit of
(\ref{eq:thirtyseven}) and (\ref{eq:thirtyeight}), when the heat
capacities of the Bose and Fermi gases become equal. Nevertheless,
for most real-world 2D systems, above the characteristic
temperature $T_C$, which marks the onset of Bose-Einstein
condensation in the two-dimensional ideal Bose gas,
${\pi\alpha\varepsilon_{C}}/ {Nk_{B}T}$ is very small and so, even
for finite gases, the thermal capacities of the Bose and Fermi
systems are approximately equal above the condensation
temperature.

In Fig. 3, we plot the reduced heat
capacities (${C_{V}}/{Nk_{B}T^{*}}$) for the Fermi and Bose gases against
the reduced temperature $T^{*}$(where $T^{*}={k_{B}T}/{\varepsilon_{C}}$) for varying
system size $N$, using the full expression for both $C_{V,Bose}$ and $C_{V,Fermi}$. As the heat capacity is approximately linear close to $T=0$ in the thermodynamic limit, we would expect the plots for the 2D Fermi gas to tend to a horizontal line as $T\rightarrow 0$. This limiting value will be at the critical value when $T^{*}=0$, i.e. 
\begin{eqnarray}
\frac{C_{V,Fermi}}{Nk_{B}T^{*}}=2 g_{2} \left(1\right) =2 \zeta\left(2\right) = \frac{\pi^{2}}{3}%
\label{eq:forty}.
\end{eqnarray}
In contrast, for the finite 2D Bose gas, we expect to see a deviation from the above mentioned low-temperature behaviour. In particular, we expect that the reduced heat capacity (${C_{V}}/{Nk_{B}T^{*}}$) will fall to 0 following the formation of the condensate. Our plots indicate distinct differences between the reduced heat capacities for the
Bose and Fermi gases, particularly at very low $T$ in the vicinity
of and below the condensation temperature $T_C$. Then,
$C_{V,Bose}$ is less than $C_{V,Fermi}$ and starts to deviate
significantly around $T_C$, before rapidly going to zero as we
approach $T=0$. This is in marked contrast to the 3D Bose gas,
where the conventional heat capacity $C_{V}/Nk_{B}$ peaks at some finite $T$. In the
two-dimensional case, however, the heat capacity $C_{V}/Nk_{B}$ is a smooth
function of $T$, in keeping with the size-dependent nature of the
condensation.  Here, instead, we see a \textquotedblleft
bump\textquotedblright in the plot for the reduced heat capacity ${C_{V}}/{Nk_{B}T^{*}}$ of the 2D Bose gas, indicating the presence of a condensation. Further, as \textit{N} increases, the behaviour tends to that of the 2D Fermi gas as the condensation forms at lower temperatures, and consequently, the fall-off of the reduced heat capacity occurs at lower temperatures. Indeed, with $N=500000$, the two plots of the Fermi and Bose gases are virtually indistinguishable. This is
significant as conventionally, the heat capacity $C_{V}/Nk_{B}$ is used
experimentally to determine if a Bose-Einstein condensation has
occurred. However, as can be seen from above, the heat capacity
does not necessarily peak for a weak transition which is dependent
on the system size, and we have to examine alternate expressions to determine the presence of a condensation.

\section{\label{Conclusion}Discussion and Conclusion}

In this paper, we set out to discuss the two-dimensional ideal
Bose and Fermi gases in the absence of any trapping potentials,
and both in the thermodynamic limit and the finite case. In
particular, we were interested in May's Theorem, and its breakdown
for finite systems.

As can be seen above in Sections \ref{The Finite Gas} and
\ref{Heat capacities}, the 2D finite Fermi system displays the
same behaviour as in the thermodynamic limit but the Bose gas
deviates significantly for a finite system and undergoes a
pseudo-Bose-Einstein condensation. It is not a \textquotedblleft
true\textquotedblright condensation as for the 3D Bose gas, where
there is a distinct phase transition in the thermodynamic limit
and a peak in the heat capacity. Rather, here we have a
characteristic condensation temperature $T_C$ that is dependent on
the system size and tends to zero in the thermodynamic limit.
Moreover, though these is no heat capacity peak here, it still
deviates significantly from the Fermi heat capacity below $T_C$
with the deviations tending to zero as the system size and/or the
temperature increase. In other words, we have a peak in the
reduced heat capacity, as shown in Fig. 3. This raises the
question of what exactly is defined as a phase transition. The
two-dimensional Bose gas undergoes a significant change at low
temperatures but only in the finite case. Thus, it is not what may
be termed a \textquotedblleft conventional\textquotedblright phase
transition as for the three-dimensional Bose gas, since it ceases
to exist in the thermodynamic limit. However, from an experimental
perspective, a system is never truly in the thermodynamic limit,
and so, what we have termed a \textquotedblleft
pseudo-condensation\textquotedblright is expected. This also,
therefore, casts doubt on the use of the heat capacity as a means
of identifying phase transitions, as discussed above.

As a final note, we would like to consider some of the outstanding
problems. The above analysis is a simple one and deals with only
the \textit{ideal} two-dimensional gas. There is still
considerable debate whether such behaviour is to be found for an
\textit{interacting} two-dimensional gas. Some authors\cite{petrov2000} have claimed that there is a true condensation for the finite interacting 2D system at sufficiently low temperatures as the coherence length becomes larger than the condensate size. Others, however, have claimed that any phase transition would disappear in the presence of particle interactions\cite{mullin97}, as predicted by the Hohenberg theorem. Clearly, the presence of
interactions will modify our results greatly. As yet, however,
there is no indication if there is any correspondence between the
two-dimensional Fermi and Bose gases, when they are interacting,
and further, if a pseudo-condensation is to be found for the
finite Bose gas. Further, similar analogues exist in other
dimensions for ideal gases in a trapping
potential\cite{ketterle96,mullin97}. It would be interesting to
debate whether there are deeper implications due to the Fermi-Bose
correspondence, and where such an situation occurs, it is a
special unique case. This is an area of active work, with new
experimental innovations making the resolution of the above
problems and the elucidation of two-dimensional gases only a
matter of time now.

\begin{figure}
\includegraphics[width=4.5in]{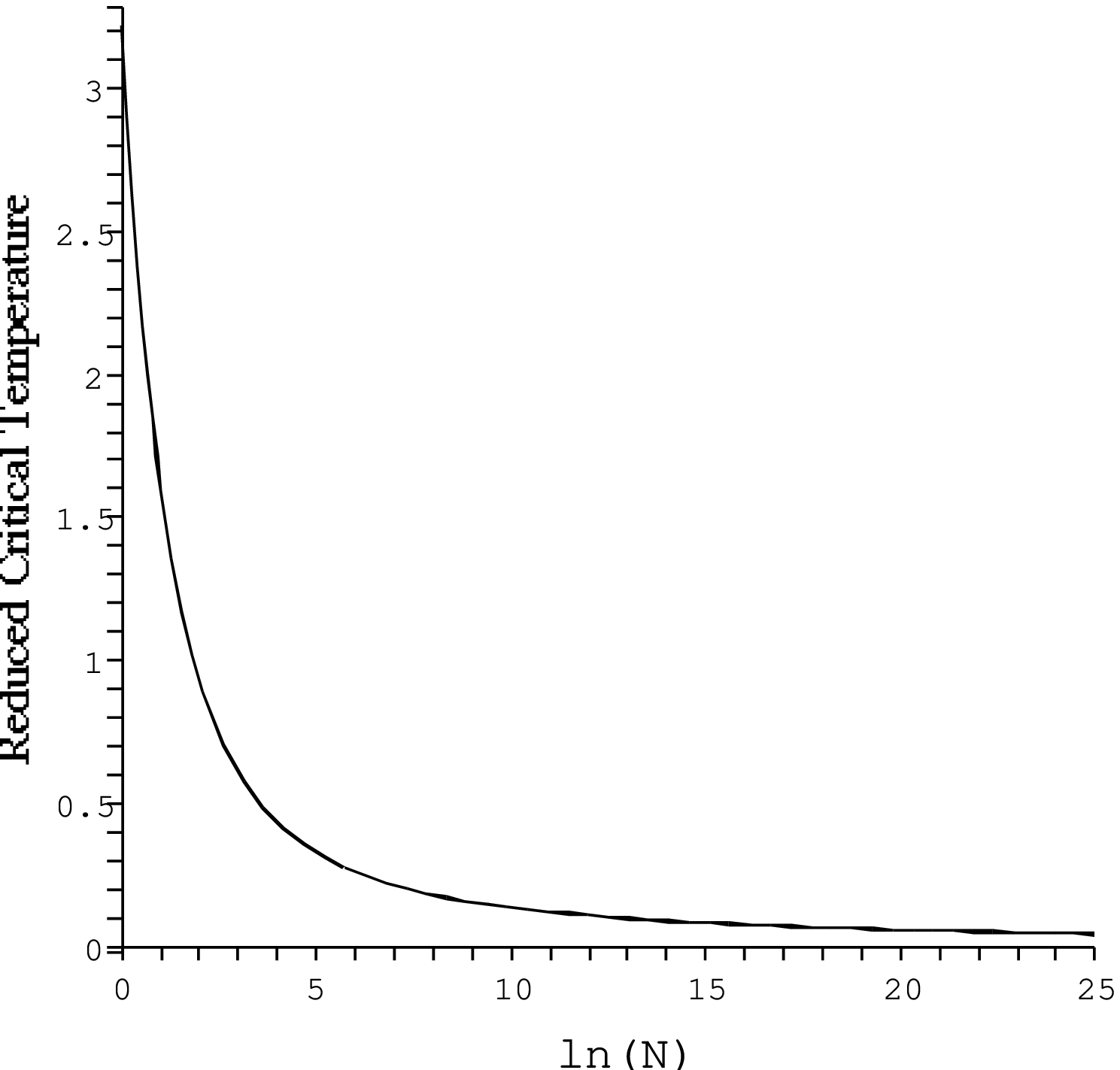}
\caption{Plot of reduced characteristic condensation temperature $k_{B}T_{C}/\varepsilon_{C}$ against number of particles $N$ in the system. [N.B. Spin degeneracy $\alpha=1$]}
\end{figure}

\begin{figure}
\includegraphics[width=4.5in]{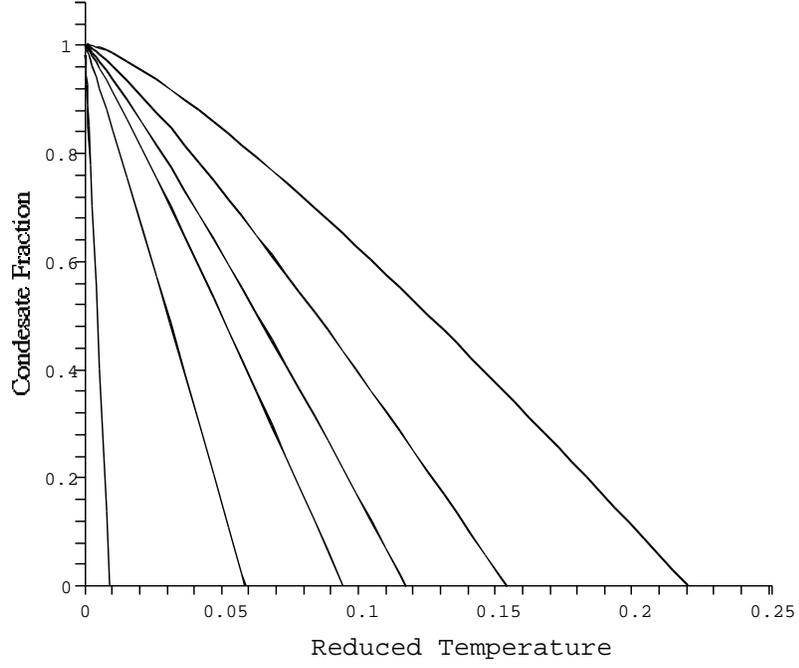}
\caption{Plot of condensate fraction $N_{0}/N$ against reduced temperature $k_{B}T/\varepsilon_{C}$ for (from right to left) $N$ = $10^{3}$, $10^{4}$, $10^{5}$, $10^{6}$, $10^{9}$, $10^{50}$. [$\alpha=1$]}
\end{figure}

\clearpage
\begin{figure}
\includegraphics[width=4in, trim=0 105 0 0]{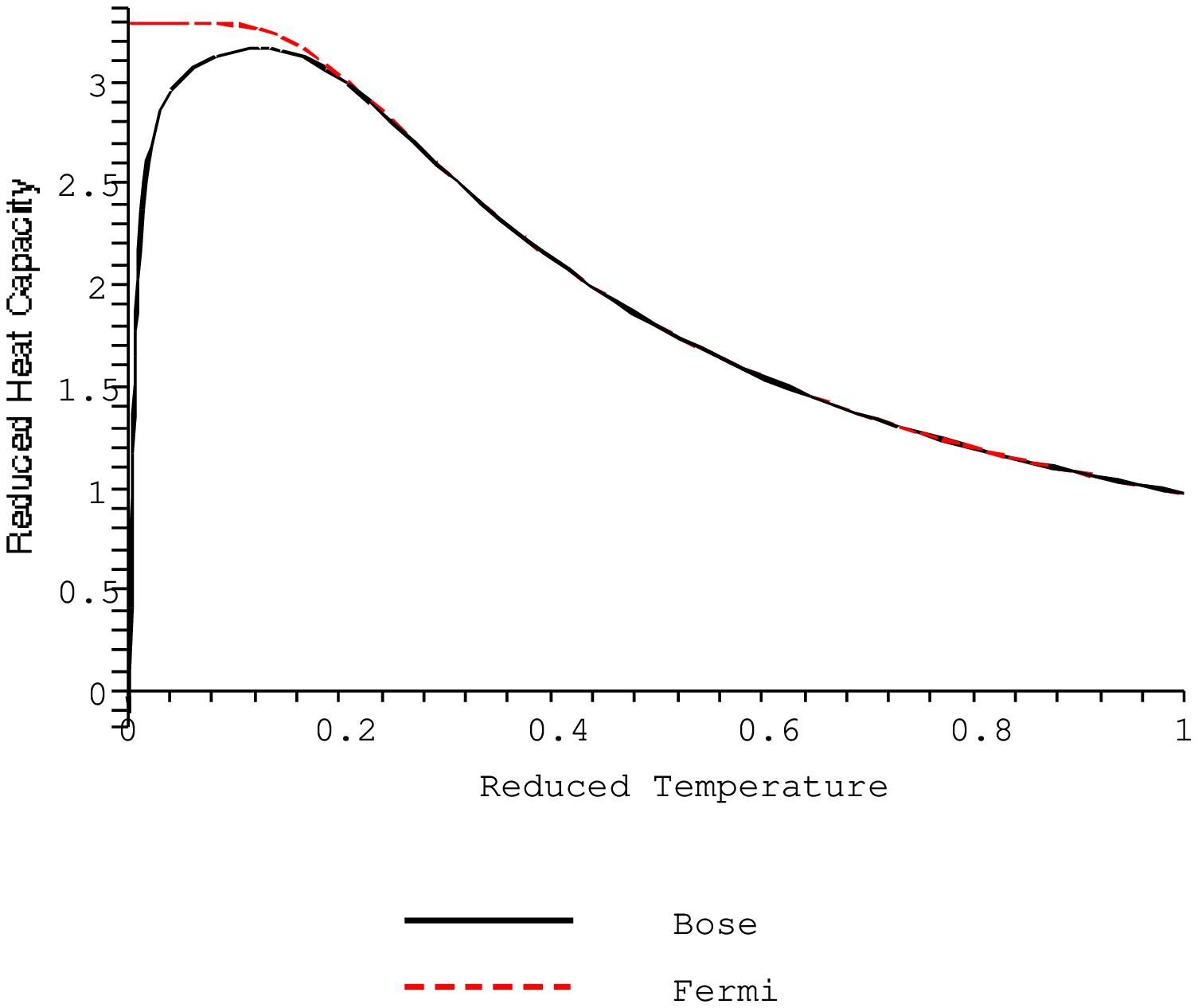}
\end{figure}

\begin{figure}
\includegraphics[width=4in, trim=0 90 0 0]{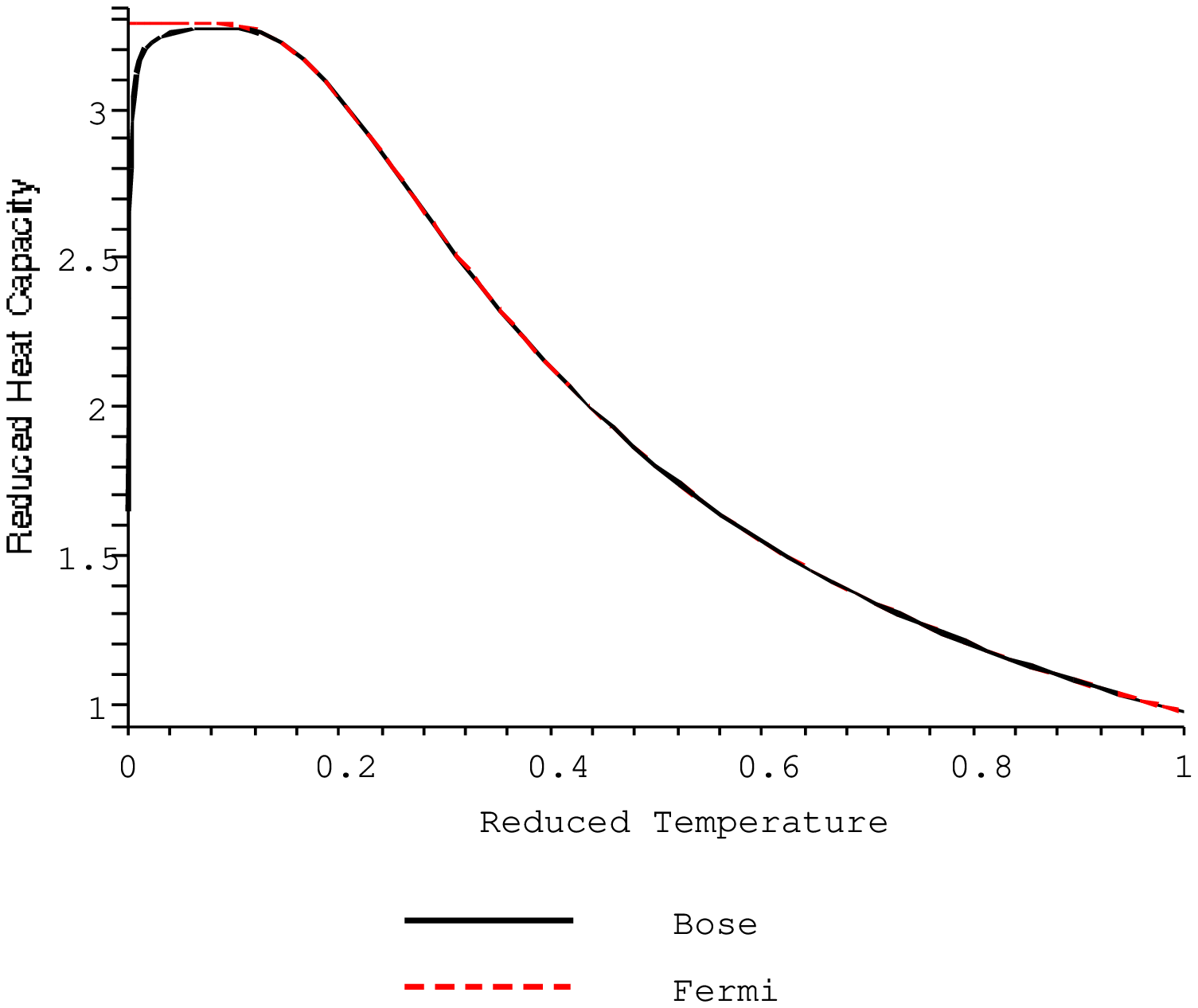}
\end{figure}

\begin{figure}
\includegraphics[width=4in, trim=0 105 0 0]{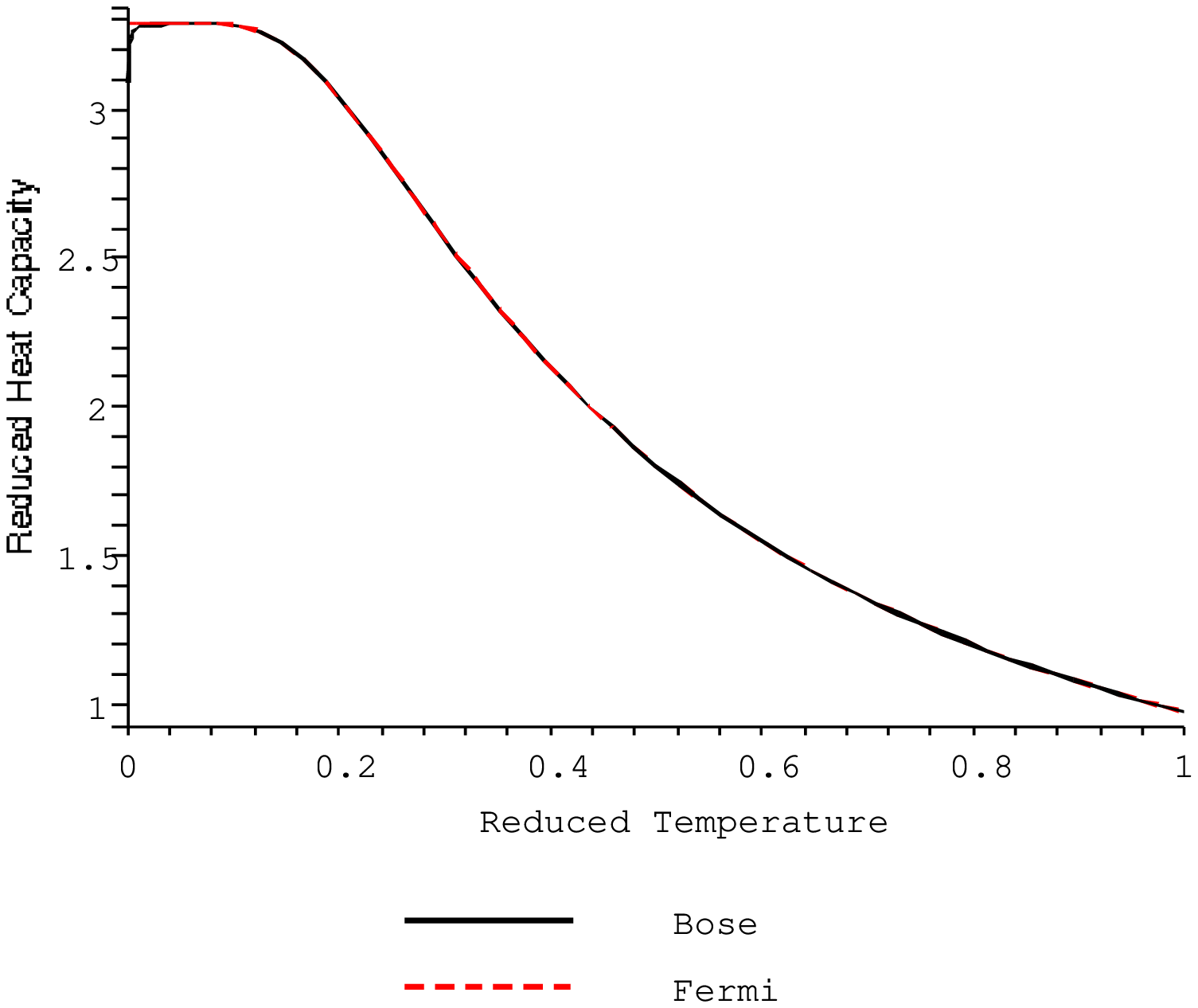}
\end{figure}

\begin{figure}
\includegraphics[width=4in]{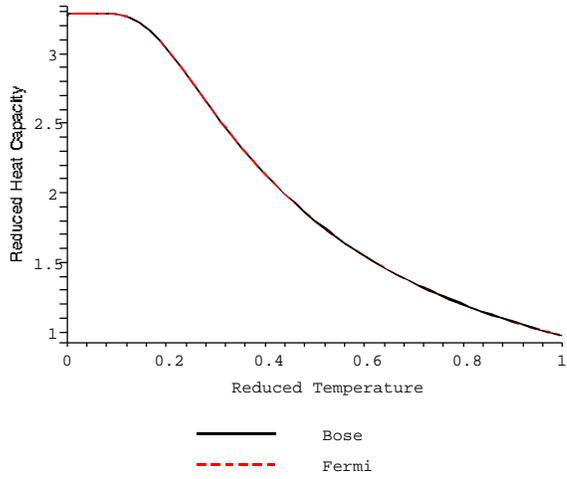}
\caption{Plot of difference in the reduced heat capacities $C_{V}/Nk_{B}T^{*}$ of the Fermi and Bose gases against reduced temperature $T^{*}=k_{B}T/\varepsilon_{C}$ for (from top to bottom) $N$ = : 500, 5000, 50000, 500000. [$\alpha=1$]}
\end{figure}


\begin{thebibliography}{33}
\expandafter\ifx\csname natexlab\endcsname\relax\def\natexlab#1{#1}\fi
\expandafter\ifx\csname bibnamefont\endcsname\relax
  \def\bibnamefont#1{#1}\fi
\expandafter\ifx\csname bibfnamefont\endcsname\relax
  \def\bibfnamefont#1{#1}\fi
\expandafter\ifx\csname citenamefont\endcsname\relax
  \def\citenamefont#1{#1}\fi
\expandafter\ifx\csname url\endcsname\relax
  \def\url#1{\texttt{#1}}\fi
\expandafter\ifx\csname urlprefix\endcsname\relax\def\urlprefix{URL }\fi
\providecommand{\bibinfo}[2]{#2}
\providecommand{\eprint}[2][]{\url{#2}}

\bibitem[{\citenamefont{Huang}(1987)}]{huang87}
\bibinfo{author}{\bibfnamefont{K.}~\bibnamefont{Huang}},
  \emph{\bibinfo{title}{Statistical Mechanics}} (\bibinfo{publisher}{John Wiley
  and Sons}, \bibinfo{address}{New York}, \bibinfo{year}{1987}),
  \bibinfo{edition}{2nd} ed.

\bibitem[{\citenamefont{Pathria}(1996)}]{pathria96}
\bibinfo{author}{\bibfnamefont{R.~K.} \bibnamefont{Pathria}},
  \emph{\bibinfo{title}{Statistical Mechanics}}
  (\bibinfo{publisher}{Butterworth-Heinemann}, \bibinfo{address}{Oxford},
  \bibinfo{year}{1996}), \bibinfo{edition}{2nd} ed.

\bibitem[{\citenamefont{Lusher et~al.}(1991)\citenamefont{Lusher, Saunders, and
  Cowan}}]{lusher91}
\bibinfo{author}{\bibfnamefont{C.~P.} \bibnamefont{Lusher}},
  \bibinfo{author}{\bibfnamefont{J.}~\bibnamefont{Saunders}}, \bibnamefont{and}
  \bibinfo{author}{\bibfnamefont{B.~P.} \bibnamefont{Cowan}},
  \bibinfo{journal}{Physical Review Letters} \textbf{\bibinfo{volume}{67}},
  \bibinfo{pages}{2199} (\bibinfo{year}{1991}).

\bibitem[{\citenamefont{Siqueira et~al.}(1993)\citenamefont{Siqueira, Lusher,
  Cowan, and Saunders}}]{siqueira93}
\bibinfo{author}{\bibfnamefont{M.}~\bibnamefont{Siqueira}},
  \bibinfo{author}{\bibfnamefont{C.~P.} \bibnamefont{Lusher}},
  \bibinfo{author}{\bibfnamefont{B.~P.} \bibnamefont{Cowan}}, \bibnamefont{and}
  \bibinfo{author}{\bibfnamefont{J.}~\bibnamefont{Saunders}},
  \bibinfo{journal}{Physical Review Letters} \textbf{\bibinfo{volume}{71}},
  \bibinfo{pages}{1407} (\bibinfo{year}{1993}).

\bibitem[{\citenamefont{Casey et~al.}(1998)\citenamefont{Casey, Patel, Nyeki,
  Cowan, and Saunders}}]{casey98}
\bibinfo{author}{\bibfnamefont{A.}~\bibnamefont{Casey}},
  \bibinfo{author}{\bibfnamefont{H.}~\bibnamefont{Patel}},
  \bibinfo{author}{\bibfnamefont{J.}~\bibnamefont{Nyeki}},
  \bibinfo{author}{\bibfnamefont{B.~P.} \bibnamefont{Cowan}}, \bibnamefont{and}
  \bibinfo{author}{\bibfnamefont{J.}~\bibnamefont{Saunders}},
  \bibinfo{journal}{Journal of Low Temperature Physics}
  \textbf{\bibinfo{volume}{113}}, \bibinfo{pages}{293} (\bibinfo{year}{1998}).

\bibitem[{\citenamefont{Ray et~al.}(1998)\citenamefont{Ray, Nyeki, Cowan, and
  Saunders}}]{ray98}
\bibinfo{author}{\bibfnamefont{R.}~\bibnamefont{Ray}},
  \bibinfo{author}{\bibfnamefont{J.}~\bibnamefont{Nyeki}},
  \bibinfo{author}{\bibfnamefont{B.~P.} \bibnamefont{Cowan}}, \bibnamefont{and}
  \bibinfo{author}{\bibfnamefont{J.}~\bibnamefont{Saunders}},
  \bibinfo{journal}{Physical Review Letters} \textbf{\bibinfo{volume}{81}},
  \bibinfo{pages}{152} (\bibinfo{year}{1998}).

\bibitem[{\citenamefont{Schakel}(1998)}]{schakel98}
\bibinfo{author}{\bibfnamefont{A.~M.~J.} \bibnamefont{Schakel}}
  (\bibinfo{year}{1998}), \eprint{cond-mat/990867}.

\bibitem[{\citenamefont{Anderson et~al.}(1995)\citenamefont{Anderson, Ensher,
  Matthews, Wieman, and Cornell}}]{anderson95}
\bibinfo{author}{\bibfnamefont{M.~H.} \bibnamefont{Anderson}},
  \bibinfo{author}{\bibfnamefont{J.~R.} \bibnamefont{Ensher}},
  \bibinfo{author}{\bibfnamefont{M.~R.} \bibnamefont{Matthews}},
  \bibinfo{author}{\bibfnamefont{C.~E.} \bibnamefont{Wieman}},
  \bibnamefont{and} \bibinfo{author}{\bibfnamefont{E.~A.}
  \bibnamefont{Cornell}}, \bibinfo{journal}{Science}
  \textbf{\bibinfo{volume}{269}}, \bibinfo{pages}{198} (\bibinfo{year}{1995}).

\bibitem[{\citenamefont{Bradley et~al.}(1997)\citenamefont{Bradley, Sackett,
  and Hulet}}]{bradley97}
\bibinfo{author}{\bibfnamefont{C.~C.} \bibnamefont{Bradley}},
  \bibinfo{author}{\bibfnamefont{C.~A.} \bibnamefont{Sackett}},
  \bibnamefont{and} \bibinfo{author}{\bibfnamefont{R.~G.} \bibnamefont{Hulet}},
  \bibinfo{journal}{Physical Review Letters} \textbf{\bibinfo{volume}{78}},
  \bibinfo{pages}{985} (\bibinfo{year}{1997}).

\bibitem[{\citenamefont{Davis et~al.}(1995)\citenamefont{Davis, Mewes, Andrews,
  van Druten, Durfee, Kurn, and Ketterle}}]{davis95}
\bibinfo{author}{\bibfnamefont{K.~B.} \bibnamefont{Davis}},
  \bibinfo{author}{\bibfnamefont{M.-O.} \bibnamefont{Mewes}},
  \bibinfo{author}{\bibfnamefont{M.~R.} \bibnamefont{Andrews}},
  \bibinfo{author}{\bibfnamefont{N.~J.} \bibnamefont{van Druten}},
  \bibinfo{author}{\bibfnamefont{D.~S.} \bibnamefont{Durfee}},
  \bibinfo{author}{\bibfnamefont{D.~M.} \bibnamefont{Kurn}}, \bibnamefont{and}
  \bibinfo{author}{\bibfnamefont{W.}~\bibnamefont{Ketterle}},
  \bibinfo{journal}{Physical Review Letters} \textbf{\bibinfo{volume}{75}},
  \bibinfo{pages}{3969} (\bibinfo{year}{1995}).

\bibitem[{\citenamefont{DeMarco and Jin}(1999)}]{demarco99}
\bibinfo{author}{\bibfnamefont{B.}~\bibnamefont{DeMarco}} \bibnamefont{and}
  \bibinfo{author}{\bibfnamefont{D.~S.} \bibnamefont{Jin}},
  \bibinfo{journal}{Science} \textbf{\bibinfo{volume}{285}},
  \bibinfo{pages}{903} (\bibinfo{year}{1999}).

\bibitem[{\citenamefont{May}(1964)}]{may64}
\bibinfo{author}{\bibfnamefont{R.~M.} \bibnamefont{May}},
  \bibinfo{journal}{Physical Review} \textbf{\bibinfo{volume}{135}},
  \bibinfo{pages}{A1515} (\bibinfo{year}{1964}).

\bibitem[{\citenamefont{Mullin and Fernandez}(2003)}]{mullin2003}
\bibinfo{author}{\bibfnamefont{W.~J.} \bibnamefont{Mullin}} \bibnamefont{and}
  \bibinfo{author}{\bibfnamefont{J.~P.} \bibnamefont{Fernandez}},
  \bibinfo{journal}{American Journal of Physics} \textbf{\bibinfo{volume}{71}},
  \bibinfo{pages}{661} (\bibinfo{year}{2003}).

\bibitem[{\citenamefont{Toda et~al.}(1983)\citenamefont{Toda, Kubo, and
  Saito}}]{toda83}
\bibinfo{author}{\bibfnamefont{M.}~\bibnamefont{Toda}},
  \bibinfo{author}{\bibfnamefont{R.}~\bibnamefont{Kubo}}, \bibnamefont{and}
  \bibinfo{author}{\bibfnamefont{N.}~\bibnamefont{Saito}},
  \emph{\bibinfo{title}{Statistical Physics I}} (\bibinfo{publisher}{Springer
  Verlag}, \bibinfo{address}{Berlin}, \bibinfo{year}{1983}).

\bibitem[{\citenamefont{Aldrovandi}(1992)}]{aldrovandi92}
\bibinfo{author}{\bibfnamefont{R.}~\bibnamefont{Aldrovandi}},
  \bibinfo{journal}{Fortschritte der Physik} \textbf{\bibinfo{volume}{40}},
  \bibinfo{pages}{631} (\bibinfo{year}{1992}).

\bibitem[{\citenamefont{Viefers et~al.}(1995)\citenamefont{Viefers, Ravndal,
  and Haugset}}]{viefers95}
\bibinfo{author}{\bibfnamefont{S.}~\bibnamefont{Viefers}},
  \bibinfo{author}{\bibfnamefont{F.}~\bibnamefont{Ravndal}}, \bibnamefont{and}
  \bibinfo{author}{\bibfnamefont{T.}~\bibnamefont{Haugset}},
  \bibinfo{journal}{American Journal of Physics} \textbf{\bibinfo{volume}{63}},
  \bibinfo{pages}{369} (\bibinfo{year}{1995}).

\bibitem[{\citenamefont{Lee}(1997)}]{lee97}
\bibinfo{author}{\bibfnamefont{M.}~\bibnamefont{Lee}},
  \bibinfo{journal}{Physical Review E} \textbf{\bibinfo{volume}{55}},
  \bibinfo{pages}{1518} (\bibinfo{year}{1997}).

\bibitem[{\citenamefont{Abramowitz and Stegun}(1964)}]{abram64}
\bibinfo{author}{\bibfnamefont{M.}~\bibnamefont{Abramowitz}} \bibnamefont{and}
  \bibinfo{author}{\bibfnamefont{I.~A.} \bibnamefont{Stegun}},
  \emph{\bibinfo{title}{Handbook of Mathematical Functions}}
  (\bibinfo{publisher}{National Bureau of Standards},
  \bibinfo{address}{Washington DC}, \bibinfo{year}{1964}).

\bibitem[{\citenamefont{Lewin}(1958)}]{lewin58}
\bibinfo{author}{\bibfnamefont{L.}~\bibnamefont{Lewin}},
  \emph{\bibinfo{title}{Dilogarithms and Associated Functions}}
  (\bibinfo{publisher}{McDonald}, \bibinfo{address}{London},
  \bibinfo{year}{1958}).

\bibitem[{\citenamefont{Osborne}(1949)}]{osborne49}
\bibinfo{author}{\bibfnamefont{M.~F.~M.} \bibnamefont{Osborne}},
  \bibinfo{journal}{Physical Review} \textbf{\bibinfo{volume}{76}},
  \bibinfo{pages}{396} (\bibinfo{year}{1949}).

\bibitem[{\citenamefont{Ziman}(1953)}]{ziman53}
\bibinfo{author}{\bibfnamefont{J.~M.} \bibnamefont{Ziman}},
  \bibinfo{journal}{Philosophical Magazine} \textbf{\bibinfo{volume}{44}},
  \bibinfo{pages}{548} (\bibinfo{year}{1953}).

\bibitem[{\citenamefont{Mills}(1964)}]{mills64}
\bibinfo{author}{\bibfnamefont{D.~L.} \bibnamefont{Mills}},
  \bibinfo{journal}{Physical Review} \textbf{\bibinfo{volume}{134}},
  \bibinfo{pages}{A306} (\bibinfo{year}{1964}).

\bibitem[{\citenamefont{Goble and Trainor}(1967)}]{goble67}
\bibinfo{author}{\bibfnamefont{D.~F.} \bibnamefont{Goble}} \bibnamefont{and}
  \bibinfo{author}{\bibfnamefont{L.~E.~H.} \bibnamefont{Trainor}},
  \bibinfo{journal}{Physical Review} \textbf{\bibinfo{volume}{157}},
  \bibinfo{pages}{167} (\bibinfo{year}{1967}).

\bibitem[{\citenamefont{Chester}(1968)}]{chester68}
\bibinfo{author}{\bibfnamefont{C.~V.} \bibnamefont{Chester}},
  \emph{\bibinfo{title}{Lectures in Theoretical Physics}}, vol.
  \bibinfo{volume}{11B} (\bibinfo{publisher}{Gordon and Breach, Scientific
  Publishers Inc.}, \bibinfo{address}{New York}, \bibinfo{year}{1968}).

\bibitem[{\citenamefont{Krueger}(1968)}]{krueger68}
\bibinfo{author}{\bibfnamefont{D.~A.} \bibnamefont{Krueger}},
  \bibinfo{journal}{Physical Review} \textbf{\bibinfo{volume}{172}},
  \bibinfo{pages}{211} (\bibinfo{year}{1968}).

\bibitem[{\citenamefont{Irnry}(1969)}]{irnry69}
\bibinfo{author}{\bibfnamefont{Y.}~\bibnamefont{Irnry}},
  \bibinfo{journal}{Annals of Physics} \textbf{\bibinfo{volume}{51}},
  \bibinfo{pages}{1} (\bibinfo{year}{1969}).

\bibitem[{\citenamefont{Barber}(1983)}]{barber83}
\bibinfo{author}{\bibfnamefont{M.~N.} \bibnamefont{Barber}},
  \emph{\bibinfo{title}{Phase Transitions and Critical Phenomena}},
  vol.~\bibinfo{volume}{8} (\bibinfo{publisher}{Academic Press},
  \bibinfo{address}{London}, \bibinfo{year}{1983}).

\bibitem[{\citenamefont{Grossmann and Holthaus}(1995)}]{grossmann95}
\bibinfo{author}{\bibfnamefont{S.}~\bibnamefont{Grossmann}} \bibnamefont{and}
  \bibinfo{author}{\bibfnamefont{M.}~\bibnamefont{Holthaus}},
  \bibinfo{journal}{Zeitschrift fur Physik B} \textbf{\bibinfo{volume}{97}},
  \bibinfo{pages}{319} (\bibinfo{year}{1995}).

\bibitem[{\citenamefont{Pathria}(1998)}]{pathria98}
\bibinfo{author}{\bibfnamefont{R.~K.} \bibnamefont{Pathria}},
  \bibinfo{journal}{Physical Review E} \textbf{\bibinfo{volume}{57}},
  \bibinfo{pages}{2697} (\bibinfo{year}{1998}).

\bibitem[{\citenamefont{Corless et~al.}(1996)\citenamefont{Corless, Gonnet,
  Hare, Jeffrey, and Knuth}}]{corless96}
\bibinfo{author}{\bibfnamefont{R.~M.} \bibnamefont{Corless}},
  \bibinfo{author}{\bibfnamefont{G.~H.} \bibnamefont{Gonnet}},
  \bibinfo{author}{\bibfnamefont{D.~E.~G.} \bibnamefont{Hare}},
  \bibinfo{author}{\bibfnamefont{D.~J.} \bibnamefont{Jeffrey}},
  \bibnamefont{and} \bibinfo{author}{\bibfnamefont{D.~E.} \bibnamefont{Knuth}},
  \bibinfo{journal}{Advances in Computational Mathematics}
  \textbf{\bibinfo{volume}{5}}, \bibinfo{pages}{329} (\bibinfo{year}{1996}).

\bibitem[{\citenamefont{Petrov et~al.}(2000)\citenamefont{Petrov, Holzmann, and
  Shlyapnikov}}]{petrov2000}
\bibinfo{author}{\bibfnamefont{D.~S.} \bibnamefont{Petrov}},
  \bibinfo{author}{\bibfnamefont{M.}~\bibnamefont{Holzmann}}, \bibnamefont{and}
  \bibinfo{author}{\bibfnamefont{G.~V.} \bibnamefont{Shlyapnikov}},
  \bibinfo{journal}{Physical Review Letters} \textbf{\bibinfo{volume}{84}},
  \bibinfo{pages}{2551} (\bibinfo{year}{2000}).

\bibitem[{\citenamefont{Mullin}(1997)}]{mullin97}
\bibinfo{author}{\bibfnamefont{W.~J.} \bibnamefont{Mullin}},
  \bibinfo{journal}{Journal of Low Temperature Physics}
  \textbf{\bibinfo{volume}{106}}, \bibinfo{pages}{615} (\bibinfo{year}{1997}).

\bibitem[{\citenamefont{Ketterle and van Druten}(1996)}]{ketterle96}
\bibinfo{author}{\bibfnamefont{W.}~\bibnamefont{Ketterle}} \bibnamefont{and}
  \bibinfo{author}{\bibfnamefont{N.~J.} \bibnamefont{van Druten}},
  \bibinfo{journal}{Physical Review A} \textbf{\bibinfo{volume}{54}},
  \bibinfo{pages}{656} (\bibinfo{year}{1996}).

\end{thebibliography}
\end{document}